\documentclass[journal]{IEEEtran}
\ifCLASSINFOpdf
\else
\fi
\usepackage{amsmath}
\usepackage{amssymb}
\usepackage{graphicx}
\usepackage{hyperref}
\usepackage{booktabs}
\usepackage{multirow}
\usepackage{makecell}
\usepackage{cite}
\usepackage{xcolor}
\makeatletter
\renewcommand\subsubsection{\@startsection{subsubsection}{3}{\parindent}
  {0.1\baselineskip} 
  {0.1\baselineskip} 
  {\normalfont\normalsize\itshape}} 
\makeatother
\begin{document}
\title{Synchronization-Safe Dynamic Microgrid Formation for DER-Led Distribution System Restoration With Constraint-Aware Graph Learning}

\author{Cong~Bai,~\IEEEmembership{Student~Member,~IEEE,} Salish~Maharjan,~\IEEEmembership{Senior~Member,~IEEE,}
Yunyi~Li,~\IEEEmembership{Member,~IEEE,}
Wenlong~Shi,~\IEEEmembership{Member,~IEEE,}
and~Zhaoyu~Wang,~\IEEEmembership{Fellow,~IEEE}
\thanks{Cong Bai, Salish Maharjan, Yunyi Li, Wenlong Shi, and Zhaoyu Wang are with the Department of Electrical and Computer Engineering, Iowa State University, Ames, IA 50011 USA (e-mail: congbai@iastate.edu; salish@iastate.edu; liyunyi@iastate.edu; wshi5@iastate.edu; wzy@iastate.edu).}
\vspace{-2.5em}}
\maketitle
\begin{abstract}
Prolonged blackouts in distribution systems (DSs) with high penetration of distributed energy resources (DERs) necessitate novel restoration strategies to rapidly restore loads. However, the resulting complex optimization problem significantly limits scalability. This paper proposes a synchronization-safe dynamic microgrid (MG) formation (SSDMGF)-enabled restoration framework, in which a constraint-aware graph learning approach is developed to enhance solution efficiency. To characterize the restoration status of systems with evolving boundaries, the concepts of system mode and system class are defined. To ensure synchronization safety during restoration, the transitions of system mode and class for dynamically formed MGs are explicitly restricted. To further accelerate the solution process, a constraint-aware spatio-temporal graph convolutional network (STGCN) is designed to partially generate high-quality warm-start solutions, where synchronization-related constraints are embedded into a differentiable feasibility-resolving layer based on the straight-through estimator (STE). Case studies on a modified IEEE 123-node feeder validate that the proposed method ensures synchronization-safe MG formation and improves restoration performance. Meanwhile, the proposed acceleration framework achieves significant computational speed-ups without compromising final optimality.
\end{abstract}
\begin{IEEEkeywords}
Black-start, dynamic microgrids, restoration, synchronization safety, graph learning.
\end{IEEEkeywords}
\IEEEpeerreviewmaketitle
\section{Introduction}
\IEEEPARstart{E}{xtended} outage durations in distribution systems (DSs), often caused by the loss of power supply from upstream transmission grids (TGs) due to frequent natural disasters, necessitate advanced restoration strategies to black-start (BS) the system and promptly restore load demand~\cite{JahidHasanRifat2025}. With the increasing penetration of distributed energy resources (DERs), reliance on centralized generation for fast and stable recovery is becoming less effective. In this context, microgrids (MGs), enabled by grid-forming inverter (GFMI)-based DERs and supported by grid-following inverter (GFLI)-based DERs, offer a flexible and decentralized solution, allowing localized and rapid restoration of loads~\cite{Wang2015, Wang2016}.

Unlike conventional centralized BS approaches, DER-led DS restoration is initiated by self-starting GFMIs, followed by the progressive formation of MGs that expand their boundaries to energize additional loads. Therefore, the dynamic formation of MGs plays a critical role in fully exploiting the capabilities of installed DERs to achieve faster and more extensive load restoration. Meanwhile, ensuring secure operation during and after MG formation is essential to maintain system stability, particularly when synchronizing multiple MGs with evolving boundaries. Although advanced control and optimization techniques have been developed to improve restoration performance, they often increase computational complexity. Consequently, efficient acceleration techniques are required to enable practical implementation.

Existing model-based restoration strategies for DSs can be broadly categorized into three groups. First, traditional centralized restoration methods rely on coordinated control from the upstream TG and system-wide optimization to determine switching and repair actions, as seen in~\cite{Cavalcante2016, Chen2019}. However, these approaches are inherently limited by slow response and reduced flexibility under widespread outages. To improve restoration resilience, DER-led restoration methods with MG formation have been widely investigated, where decentralized and faster recovery is enabled by leveraging DERs to form islanded MGs. Representative works in~\cite{Fu2020, Li2021, Cai2021, Huang2025} formulate restoration as optimization problems that determine MG configurations and load pickup sequences under various operational constraints. Despite their advantages, these approaches typically assume fixed MG boundaries or limited coordination among MGs, restricting the full utilization of DERs. More recently, dynamic MG-enabled restoration approaches have been proposed to further enhance restoration efficiency by allowing adaptive boundary expansion and merging of multiple MGs. For example, dynamic topology reconfiguration and coordinated multi-MG restoration strategies have been developed to improve load transfer flexibility and recovery speed~\cite{Zhao2025}, while probabilistic sequential restoration frameworks based on multimaster MGs incorporate dynamic constraints and uncertainty modeling to better capture realistic system behavior~\cite{Agah2025}. Nevertheless, these methods generally do not explicitly consider synchronization safety during MG formation and interconnection, such as restricting each synchronization action to only two MGs in order to avoid additional coordination requirements and excessive inrush current, which may lead to infeasible or unstable restoration actions.

Recent advances have explored artificial intelligence (AI) techniques to enhance DS restoration. Existing AI-based methods can be broadly categorized into two groups. The first category employs decision-oriented learning approaches, such as deep reinforcement learning (DRL), to directly generate restoration actions or switching sequences. Representative works in~\cite{Huang2022, Igder2023, Vu2023} utilize single-agent or multi-agent DRL frameworks to enable distributed load restoration and dynamic MG formation. While these methods offer fast inference and adaptability, they typically lack explicit guarantees on optimality and constraint satisfaction, which may limit their reliability in practical applications. The second category focuses on learning-assisted optimization, where deep learning models are integrated with mathematical programming to accelerate the solution process. Works in~\cite{Wang2024, Wu2025, Qin2025} develop end-to-end learning frameworks, deep neural network surrogates, and physics-informed graph neural networks to improve computational efficiency in DS reconfiguration and restoration. Although these approaches enhance scalability, they generally do not explicitly enforce operational and physical constraints within the learning process, potentially resulting in infeasible or suboptimal solutions.

Despite the above advances, several critical challenges remain in enabling efficient and reliable DER-led DS restoration. First, existing dynamic MG formation strategies primarily focus on boundary expansion and coordination among MGs, but lack explicit modeling of synchronization safety during MG merging, which is essential to ensure stable system operation. Second, while AI-based methods have shown promise in improving computational efficiency, current approaches either sacrifice optimality and feasibility guarantees or fail to explicitly incorporate physical and operational constraints into the learning process. As a result, there remains a lack of a unified framework that can simultaneously ensure synchronization-safe MG formation, preserve solution feasibility and optimality, and accelerate the restoration process.

Hence, this paper proposes a synchronization-safe dynamic MG formation (SSDMGF)-enabled restoration framework, which ensures secure MG merging while improving computational efficiency through a constraint-aware graph learning layer. The main contributions of this work are summarized as follows:
\begin{itemize}
    \item A configurable synchronization structure is established to explicitly capture the coupling between BS sources and synchronizing switches (SSWs). Based on this structure, the system mode and system class are defined to characterize the restoration status of a system with evolving boundaries.
    \item An SSDMGF-enabled restoration problem is formulated for DSs, in which system mode and class transitions are dynamically restricted to ensure synchronization-safe MG merging during system energization.
    \item A constraint-aware STGCN is developed to generate high-quality partial warm starts, thereby accelerating the solution process. Synchronization-related constraints are embedded into a differentiable feasibility-resolving layer using the straight-through estimator (STE), enabling constraint-aware learning.
\end{itemize}

The remainder of this paper is organized as follows. Section~\ref{sec:ii} introduces graph-related preliminaries. Section~\ref{sec:iii} presents the mathematical formulation of the SSDMGF-enabled restoration problem. Section~\ref{sec:iv} develops the proposed constraint-aware graph learning-based acceleration framework. Section~\ref{sec:v} presents case studies to validate the effectiveness of the proposed method. Finally, Section~\ref{sec:vi} concludes the paper.
\section{Preliminary}\label{sec:ii}
Let $\mathcal{L}=\mathcal{L}^{\mathrm{LN}}\cup\mathcal{L}^{\mathrm{SW}}$ denote the set of lines in a DS, where $\mathcal{L}^{\mathrm{LN}}$ and $\mathcal{L}^{\mathrm{SW}}$ represent the sets of non-switchable lines and switchable lines, respectively. By opening switches in $\mathcal{L}^{\mathrm{SW}}$, the DS can be partitioned into multiple bus blocks, each consisting of buses interconnected through non-switchable lines in $\mathcal{L}^{\mathrm{LN}}$.

Each switchable line connects two buses. Let $\mathcal{K}$ denote the set of bus blocks, and let $\tau(\cdot)$ map each bus to its corresponding block. The induced bus-block edge set is defined accordingly
\begin{equation}
    \mathcal{E}^{\mathrm{SW}}:=\left\{\left(\tau(b),\tau(b')\right)|(b,b')\in\mathcal{L}^{\mathrm{SW}}\right\}.
\end{equation}

To unify the notation and facilitate subsequent learning-based acceleration, a backbone graph $\mathcal{G}=(\mathcal{K},\mathcal{E}^{\mathrm{SW}})$ is constructed using bus blocks as vertices and induced switchable connections as edges. This graph represents a candidate interaction structure rather than an operable system topology, capturing all potential interconnections that may arise during restoration.

During the restoration horizon, only a subset of vertices and edges in $\mathcal{G}$ is energized at each time step, forming a time-varying operational topology. This topology evolves with restoration actions and remains a subgraph of $\mathcal{G}$.

For any two distinct bus blocks, a path in $\mathcal{G}$ is defined as an ordered sequence of connected vertices
\begin{equation}
    \begin{split}
    &\mathcal{P}_{k\rightarrow k'}=(k=k_0, k_1,...,k_{N_{kk'}}=k'),\\
    &\text{s.t.}\;(k_{i-1},k_{i})\in\mathcal{E}^{\mathrm{SW}}, \forall i = 1,...,N_{kk'}.
    \end{split}
\end{equation}
where $N_{kk'}$ is the number of vertices within the $\mathcal{P}_{k\rightarrow k'}$.

The corresponding vertex and edge sets of a path are defined as $\mathcal{K}^{\mathcal{P}_{k\rightarrow k'}}$ and $\mathcal{E}^{\mathcal{P}_{k\rightarrow k'}}$. A path is said to be simple if no vertex is visited more than once.
\section{Synchronization-Safe Dynamic Microgrid Formation-Enabled Restoration Formulation}\label{sec:iii}
This section presents the mathematical formulation of the SSDMGF-enabled restoration problem. First, the system mode and system class associated with dynamic MG formation are defined based on a configurable synchronization structure. Then, the objective function is introduced, followed by the modeling of synchronization-safe MG formation. Finally, BS energization constraints and system operational constraints are presented.
\vspace{-1em}
\subsection{System Mode and Class Definitions for Dynamic MG Formation}\label{subsec:iii.A}
\subsubsection{Configurable Synchronization Structure}
During the restoration, the generation units with self-starting capability are referred as BS sources. 

Two types of BS sources are considered: GFMI-based battery energy storage systems (BESS) and the upstream TG. Let $\mathcal{K}^{\mathrm{BS}}=\mathcal{K}^{\mathrm{BESS}}\cup\mathcal{K}^{\mathrm{TG}}$ denote the set of bus blocks containing BS sources, where $\mathcal{K}^{\mathrm{BESS}}$ and $\mathcal{K}^{\mathrm{TG}}$ represent bus blocks with BESS and TG connection, respectively. All elements in $\mathcal{K}^{\mathrm{BESS}}$ are always available, while those in $\mathcal{K}^{\mathrm{TG}}$ depend on the upstream grid condition. A binary parameter $u^{\mathrm{TG}}$ characterizes TG availability, defining the active BS source set $\mathcal{K}^{\mathrm{BS}}(u^{\mathrm{TG}})$.

Let $\mathcal{L}^{\mathrm{SSW}}\subseteq\mathcal{L}^{\mathrm{SW}}$ denote the set of lines equipped with SSWs, and let $\mathcal{E}^{\mathrm{SSW}}$ be the corresponding bus-block edge set. Each $e\in\mathcal{E}^{\mathrm{SSW}}$ is modeled as a configurable device that synchronizes two BS source-equipped bus blocks. All SSWs' feasible configurations are defined as
\begin{equation}
    \Omega_{e}=\{\omega_{e,0},\omega_{e,1},...,\omega_{e,N_e}\}\subseteq\tbinom{\mathcal{K}^{\mathrm{BS}}(u^{\mathrm{TG}})}{2},
\end{equation}
where $\omega_{e,0}=\varnothing$ represents the open state and $N_e$ is the number of feasible non-null configurations. $\tbinom{\cdot}{2}$ denotes the set of all subsets of 2-element of a set. Configurations involving bus blocks in $\mathcal{K}^{\mathrm{TG}}$ are excluded when $u^{\mathrm{TG}}=0$.

For $e\in\mathcal{E}^{\mathrm{SSW}}$ and $k,k'\in\mathcal{K}^{\mathrm{BS}}, k\neq k'$, if $\omega_{e}=(k,k')\in\Omega_e$, the corresponding path satisfies
\begin{equation}
    \mathcal{K}^{\mathcal{P}_{k\rightarrow k'}}\cap\mathcal{K}^{\mathrm{BS}}=\{k,k'\}, \quad 
    \mathcal{E}^{\mathcal{P}_{k\rightarrow k'}}\cap\mathcal{E}^{\mathrm{SSW}}=\{e\}.
\end{equation}
\subsubsection{System Mode and Class of Synchronization-Induced Islands}
Given a configuration vector $\boldsymbol{\omega}=\{\omega_{e}\}_{e\in\mathcal{E}^{\mathrm{SSW}}}$ and TG availability $u^{\mathrm{TG}}$, the synchronization structure is represented as
\begin{equation}
    \tilde{\mathcal{G}}(\boldsymbol{\omega},u^{\mathrm{TG}}) = \left(\mathcal{K}^{\mathrm{BS}}(u^{\mathrm{TG}}), \mathcal{E}^{\boldsymbol{\omega}}\right), \quad 
    \mathcal{E}^{\boldsymbol{\omega}}=\textstyle\bigcup_{e\in\mathcal{E}^{\mathrm{SSW}}}\omega_{e},
\end{equation}
where $\mathcal{E}^{\boldsymbol{\omega}}$ represents the virtual edges induced by the selected SSW configurations.


The system mode is defined as the partition of BS-equipped bus blocks induced by the connected components of $\tilde{\mathcal{G}}$,
\begin{equation}
    \mu\left[\tilde{\mathcal{G}}(\boldsymbol{\omega},u^{\mathrm{TG}})\right]
    :=\mathrm{ConnComp}\left[\tilde{\mathcal{G}}(\boldsymbol{\omega},u^{\mathrm{TG}})\right],
\end{equation}
where each connected component corresponds to a synchronized island.

Let $\mathcal{W}$ denote the set of all feasible configuration vectors. The set of all possible system modes is given by
\begin{equation}
    \mathcal{M}:=\mathrm{Im}(\mu)=\left\{m:\, m=\mu\left[\tilde{\mathcal{G}}(\boldsymbol{\omega},u^{\mathrm{TG}})\right],\, \boldsymbol{\omega}\in\mathcal{W}\right\}.
\end{equation}

A given system mode may correspond to multiple synchronization realizations. For a mode $m$, the set of realizations is defined as
\begin{equation}
    \mathbb{G}_{m}:=\left\{\tilde{\mathcal{G}}(\boldsymbol{\omega},u^{\mathrm{TG}}):\, 
    \mu\left[\tilde{\mathcal{G}}(\boldsymbol{\omega},u^{\mathrm{TG}})\right]=m\right\}.
\end{equation}

The system class is defined as the number of synchronized islands,
\begin{equation}
    \kappa(m):=|m|,
\end{equation}
i.e., the cardinality of the partition induced by $m$. The set of all possible system classes is
\begin{equation}
    \mathcal{C}=\{1,2,\dots,|\mathcal{K}^{\mathrm{BS}}|\}.
\end{equation}

Multiple modes may correspond to the same class. The set of modes associated with class $c$ is defined as
\begin{equation}
    \mathcal{M}_{c}= \{m\in\mathcal{M}:\kappa(m)=c\}.
\end{equation}

This abstraction preserves synchronization-relevant topological relationships while significantly reducing the complexity of the original backbone graph.
\subsection{Objective of Black-Start (BS)}
The objective is to maximize the total restored load over the restoration horizon:
\begin{equation}\label{eq:objecivefunction}
    \max\sum_{t\in\mathcal{T}}\Delta t\left(\alpha^{\mathrm{CL}}\hspace{-0.5em}\sum_{b\in\mathcal{B}^{\mathrm{CL}}}\hspace{-0.5em}\boldsymbol{1}^{T}_{\Phi_b}\boldsymbol{p}^{\mathrm{CL}}_{b,t} \hspace{-0.2em}+ \hspace{-0.2em}\alpha^{\mathrm{NL}}\hspace{-0.5em}\sum_{b\in\mathcal{B}^{\mathrm{NL}}}\hspace{-0.5em}\boldsymbol{1}^{T}_{\Phi_b}\boldsymbol{p}^{\mathrm{NL}}_{b,t}\right),
\end{equation}
where $\Delta t$ is the time step duration. $\alpha^{\mathrm{CL}}$ and $\alpha^{\mathrm{NL}}$ are weighting coefficients for critical loads (CLs) and non-critical loads (NLs), respectively. $\mathcal{B}^{\mathrm{CL}}$ and $\mathcal{B}^{\mathrm{NL}}$ denote the corresponding bus sets. $\Phi_b$ is the phase set of bus $b$, and $\boldsymbol{1}_{\Phi_b}^{T}$ is the transpose of an all-ones column vector of dimension $|\Phi_b|$. $\boldsymbol{p}^{\mathrm{CL}}_{b,t}$ and $\boldsymbol{p}^{\mathrm{NL}}_{b,t}$ represent restored active load vectors.
\vspace{-1em}
\subsection{Formulation of SSDMGF}
\subsubsection{Model of BS Source Frequency}
For $k\in\mathcal{K}^{\mathrm{TG}}$, the frequency is fixed at the nominal value when the upstream grid is available:
\begin{equation}
    f_{k,t} = 60u^{\mathrm{TG}}_{t}.
\end{equation}

For $k\in\mathcal{K}^{\mathrm{BESS}}$, the GFMI frequency is modeled using quasi-steady-state (QSS) values with synchronization adjustments:
\begin{subequations}
\begin{align}
        f_{k,t} &= f^{\mathrm{QSS}}_{k,t} + \delta^{\mathrm{sync}}_{k,t}\Delta f^{\mathrm{sync}}_{k,t},\\
        \delta^{\mathrm{sync}}_{k,t} &= \textstyle\sum_{l\in\mathcal{L}^{\mathrm{SSW}}}\Delta u^{\mathrm{SSW}}_{l,t}\sum_{k'\in\mathcal{K}^{\mathrm{BS}},k\neq k'}u^{\mathrm{sync}}_{kk',t},
\end{align}
\end{subequations}
where $f^{\mathrm{QSS}}_{k,t}$ is the QSS frequency, $\delta^{\mathrm{sync}}_{k,t}$ indicates synchronization events, and $\Delta f^{\mathrm{sync}}_{k,t}$ denotes the resulting frequency adjustment. $u^{\mathrm{SSW}}_{l,t}$ is a binary variable indicating whether SSW $l$ is closed, and $\Delta(\cdot)$ operates on its last step value to get the increment. $u^{\mathrm{sync}}_{kk',t}$ indicates whether $k$ is synchronized with $k'$.

The detailed computation of $f^{\mathrm{QSS}}_{k,t}$ and related dynamic indices (e.g., maximum rate of change of frequency (MaxRoCoF) and frequency nadir) follows \cite{Maharjan2025}. Frequency security is enforced as
\begin{equation}\label{eq:frequencysecurity}
    \lfloor f^{\chi} \rfloor \le f_{k,t}^{\chi} \le \lceil f^{\chi} \rceil,
\end{equation}
where $\chi\in\{\varnothing,\mathrm{QSS},\mathrm{MaxRoCoF},\mathrm{nadir}\}$ and $\lfloor\cdot\rfloor$ and $\lceil\cdot\rceil$ are the lower and upper limit operators of a variable, respectively. 
\subsubsection{Model of Frequency Propagation and Synchronization}
A BS source determines the frequency of its associated bus block. Let $\mathcal{B}_k$ denote the set of buses in block $k$. For switch terminal buses $\mathcal{B}^{\mathrm{SW}}$, define $\mathcal{B}^{\mathrm{SW}}_{k} = \mathcal{B}^{\mathrm{SW}}\cap\mathcal{B}_{k}$. The following constraint enforces frequency consistency:
\begin{equation}
    f_{b,t} = f_{k,t}, \quad \forall b\in\mathcal{B}^{\mathrm{SW}}_{k}.
\end{equation}

Let $\mathcal{L}^{\mathrm{ESW}}$ denote energizing switches (ESWs). For $l=(b,b')\in\mathcal{L}^{\mathrm{ESW}}$, frequency propagation during energization is enforced by
\begin{equation}\label{eq:eswfrequency}
        f_{b',t} - (1-u^{\mathrm{ESW}}_{l,t})\lceil f \rceil \le f_{b,t} \le f_{b',t} + (1-u^{\mathrm{ESW}}_{l,t})\lceil f \rceil.
\end{equation}

For SSW $l=(b,b')$, synchronization requires frequency matching:
\begin{subequations}
\begin{align}
        f_{b,t} &\le f_{b',t} + (1 - u^{\mathrm{SSW}}_{l,t})\lceil f \rceil + \epsilon,\\
        f_{b,t} &\ge f_{b',t} - (1 - u^{\mathrm{SSW}}_{l,t})\lceil f \rceil - \epsilon.
\end{align}
\end{subequations}
where $\epsilon$ is the tolerance for synchronization between two MGs.

At the closing instant, zero active and reactive power flow is enforced to ensure voltage magnitude and phase consistency:
\begin{equation}\label{eq:sswnobranchflow}
     (\Delta u^{\mathrm{SSW}}_{l,t}-1)\boldsymbol{\chi}_{l,\max}
     \le \boldsymbol{\chi}_{l,t} \le
     (1-\Delta u^{\mathrm{SSW}}_{l,t})\boldsymbol{\chi}_{l,\max}.
\end{equation}
where for $\chi\in\{p,q\}, \boldsymbol{\chi}_{l,t}$ is the column vector storing the active/reactive power through SSW $l$. $\boldsymbol{\chi}_{l,\max}$ is the column vector storing the maximum active/reactive power for $l$.
\subsubsection{Static System Class and Mode Identification}
The system class is determined by the number of synchronized islands:
\begin{subequations}\label{eq:systemclass}
\begin{align}
    \textstyle\sum_{c\in\mathcal{C}}cu^{\mathrm{C}}_{c,t} &=s_{t},\\
    \textstyle\sum_{c\in\mathcal{C}}u^{\mathrm{C}}_{c,t} &= 1,
\end{align}
\end{subequations}
where $u^{\mathrm{C}}_{c,t}$ indicates class selection and $s_t$ is the number of slack buses.

As presented in \ref{subsec:iii.A}, the relationship between system modes and classes is given by
\begin{equation}
    \textstyle\sum_{m\in \mathcal{M}_{c}} u^{\mathrm{M}}_{m,t} = u^{\mathrm{C}}_{c,t},
\end{equation}
where $u^{\mathrm{M}}_{m,t}$ is a binary variable representing whether the system is under the mode $m$ at time $t$.

Synchronization indicators are updated based on frequency matching:
\begin{subequations}
    \begin{align}
        u^{\mathrm{sync}}_{kk',t} &\ge u^{\mathrm{sync}}_{kk',t - 1},\\
        u^{\mathrm{sync}_{-}}_{kk',t} + u^{\mathrm{sync}}_{kk',t} + u^{\mathrm{sync}_{+}}_{kk',t} &= 1,\\
        2\epsilon u^{\mathrm{sync}_{-}}_{kk',t} \hspace{-0.5em}- \epsilon u^{\mathrm{sync}}_{kk',t} - \lceil f\rceil u^{\mathrm{sync}_{+}}_{kk',t} &\le f_{k',t} \hspace{-0.2em}- f_{k,t} + \mu_{kk'},\\
        \lceil f\rceil u^{\mathrm{sync}_{-}}_{kk',t} \hspace{-0.2em}+ \epsilon u^{\mathrm{sync}}_{kk',t}\hspace{-0.2em} - 2\epsilon u^{\mathrm{sync}_{+}}_{kk',t} &\ge f_{k',t} \hspace{-0.2em}- f_{k,t} + \mu_{kk'},
    \end{align}
\end{subequations}
where $u^{\mathrm{sync}_{-/+}}_{kk',t}$ are the ancillary binary variables indicating the frequency difference between $k$ and $k'$.

If $k$ and $k'$ belong to the same component under mode $m$, synchronization is enforced as
\begin{equation}
    1 - (1 - u^{\mathrm{M}}_{m,t})\le u^{\mathrm{sync}}_{kk'} \le 1 + (1 - u^{\mathrm{M}}_{m,t}),
\end{equation}
otherwise,
\vspace{-1em}
\begin{equation}
    -(1 - u^{\mathrm{M}}_{m,t})\le u^{\mathrm{sync}}_{kk'} \le 1 - u^{\mathrm{M}}_{m,t}.
\end{equation}
\subsubsection{Synchronization-Safe System Mode Transition}
Additionally, to decrease the inrush current of the GFMI-based BESS at the synchronization moment, and avoid introducing the center controller to coordinate the synchronization process, the number of MGs merged into another MG should be restricted to one. For any three distinct BS source-equipped bus blocks $k$, $k'$, and $k''$ in $\mathcal{K}^{\mathrm{BS}}$, the above extra synchronization safety constraint is modeled as,
\begin{equation}\label{eq:systemmodetransition}
    \Delta u^{\mathrm{sync}}_{kk',t} + \Delta u^{\mathrm{sync}}_{kk'',t} - u^{\mathrm{sync}}_{k'k'',t-1} \le 1.
\end{equation}

Given the relationship between the system mode and the synchronization indicators, this synchronization safety condition can be viewed as restricting some unsafe system mode transition during the restoration.
\subsection{Formulation of BS Energization}
\subsubsection{Bus Block Energization Status}
Let $\mathcal{L}_k$ denote the set of lines within bus block $k$. The subsets of non-switchable lines and ESWs in $k$ are defined as $\mathcal{L}^{\mathrm{LN}}_{k} = \mathcal{L}^{\mathrm{LN}} \cap \mathcal{L}_{k}$ and $\mathcal{L}^{\mathrm{ESW}}_{k} = \mathcal{L}^{\mathrm{ESW}} \cap \mathcal{L}_{k}$, respectively. The energization status of $k$ is constrained by
\begin{subequations}\label{eq:segmentstatus}
    \begin{align}
    u^{\mathrm{BK}}_{k,t} &\ge u^{\mathrm{BK}}_{k,t - 1},\label{eq:segmentstatus1}\\
    u^{\mathrm{BK}}_{k,t} &= u^{\mathrm{B}}_{b,t}, \forall b\in\mathcal{B}_{k},\label{eq:segmentstatus2}\\
    u^{\mathrm{BK}}_{k,t} &= u^{\mathrm{L}}_{l,t}, \forall l\in\mathcal{L}^{\mathrm{LN}}_{k},\label{eq:segmentstatus3}\\
    u^{\mathrm{BK}}_{k,t} &\ge u^{\mathrm{ESW}}_{l,t}, \forall l\in\mathcal{L}^{\mathrm{ESW}}_{k},\label{eq:segmentstatus4}\\
    \textstyle\sum_{l\in\mathcal{L}^{\mathrm{ESW}}_{k}}\Delta u^{\mathrm{ESW}}_{l,t} &\le u^{\mathrm{BK}}_{k,t - 1}M + 1,\label{eq:segmentstatus5}
    \end{align}
\end{subequations}
where $u^{\mathrm{BK}}_{k,t}$ is a binary variable indicating whether bus block $k$ is energized at time $t$, $u^{\mathrm{B}}_{b,t}$ and $u^{\mathrm{L}}_{l,t}$ denote the energization status of bus $b$ and line $l$, respectively, and $u^{\mathrm{ESW}}_{l,t}$ indicates whether ESW $l$ is closed. Here, $M_k=\max\{0,|\mathcal{L}^{\mathrm{ESW}}_{k}|-2\}$.

Constraint \eqref{eq:segmentstatus1} enforces monotonic energization. Constraints \eqref{eq:segmentstatus2} and \eqref{eq:segmentstatus3} ensure that all buses and non-switchable lines within a bus block share the same energization status. Constraints \eqref{eq:segmentstatus4} and \eqref{eq:segmentstatus5} ensure that at most one ESW is newly closed to energize a previously de-energized bus block.
\subsubsection{Model of Switchable Lines' Action}
For each ESW $l=(b,b')\in\mathcal{L}^{\mathrm{ESW}}$, the operational constraints are
\begin{subequations}\label{eq:eswaction}
    \begin{align}
    u^{\mathrm{ESW}}_{l,t} &\le u^{\mathrm{B}}_{b,t-1} + u^{\mathrm{B}}_{b',t-1}, \label{eq:eswaction1}\\
    \Delta u^{\mathrm{ESW}}_{l,t} &\le 2 - u^{\mathrm{B}}_{b,t-1} - u^{\mathrm{B}}_{b',t-1}. \label{eq:eswaction2}
    \end{align}
\end{subequations}

These constraints allow an ESW to energize a de-energized bus block only when at least one terminal bus is already energized, while preventing its use for synchronizing two isolated MGs.

For each SSW $l=(b,b')\in\mathcal{L}^{\mathrm{SSW}}$, the operation is constrained by
\begin{subequations}\label{eq:sswaction}
    \begin{align}
    u^{\mathrm{SSW}}_{l,t} &\ge u^{\mathrm{SSW}}_{l,t-1}, \label{eq:sswaction1}\\
    2u^{\mathrm{SSW}}_{l,t} &\le u^{\mathrm{B}}_{b,t-1} + u^{\mathrm{B}}_{b',t-1}. \label{eq:sswaction2}
    \end{align}
\end{subequations}

These constraints imply that SSW actions are irreversible and that both terminals must be energized before closing.
\subsubsection{Radiality of the DS}
To maintain safe and stable operation during BS, the DS must remain radial. This is enforced by
\begin{subequations}\label{eq:radiality}
    \begin{align}
    \textstyle\sum_{l\in\mathcal{L}}u^{\mathrm{L}}_{l,t} &= \sum_{b\in\mathcal{B}}u^{\mathrm{B}}_{b,t} - s_{t},\label{eq:radiality1}\\
    s_{t} &= |\mathcal{K}^{\mathrm{BESS}}| + u^{\mathrm{TG}}_{t} - \textstyle\sum_{l\in\mathcal{L}^\mathrm{SSW}}u^{\mathrm{SSW}}_{l,t}.\label{eq:radiality2}
    \end{align}
\end{subequations}

Constraint~\eqref{eq:radiality1} enforces radiality by balancing energized buses and lines, while \eqref{eq:radiality2} defines the number of energized islands at each time step.
\subsection{Formulation of System Operation}
\subsubsection{Model of TG Output}
The TG is unavailable until the upstream fault is cleared. Let $\mathcal{B}^{\mathrm{TG}}$ denote the set of buses connected to the TG. For each $b\in\mathcal{B}^{\mathrm{TG}}$, the TG output at time $t$ satisfies
\begin{equation}
        \left(\tfrac{1}{3}u^{\mathrm{TG}}_{t}S^{\mathrm{TG}}_{b,\max}\right)^2
        \ge \max_{n\in\Phi_b}\left\{\left(p^{\mathrm{TG}}_{b,n,t}\right)^2+\left(q^{\mathrm{TG}}_{b,n,t}\right)^2\right\},
\end{equation}
where $p^{\mathrm{TG}}_{b,n,t}$ and $q^{\mathrm{TG}}_{b,n,t}$ are the active and reactive power outputs of the TG at phase $n$, and $S^{\mathrm{TG}}_{b,\max}$ is the rated apparent power limit.
\subsubsection{Model of BESS Output}
Let $\mathcal{B}^{\mathrm{BESS}}$ denote the set of buses equipped with BESSs. For each $b\in\mathcal{B}^{\mathrm{BESS}}$, the BESS output is constrained by
\begin{subequations}\label{eq:BESScontrol}
\begin{align}
    \left(\tfrac{1}{3}S^{\mathrm{BESS}}_{b,\mathrm{nom}}\right)^2
    &\ge \max_{n\in\Phi_b}\left\{\left(p^{\mathrm{BESS}}_{b,n,t}\right)^2+\left(q^{\mathrm{BESS}}_{b,n,t}\right)^2\right\},\\
    SoC_{b,t} &= SoC_{b,t-1} - \frac{\boldsymbol{1}_{\Phi_b}^{T}\boldsymbol{p}^{\mathrm{BESS}}_{b,t}\Delta t}{E^{\mathrm{BESS}}_{b,\mathrm{nom}}},\\
    \lfloor SoC \rfloor &\le SoC_{b,t} \le \lceil SoC \rceil, \label{eq:SoCconstraint}
\end{align}
\end{subequations}
where $p^{\mathrm{BESS}}_{b,n,t}$ and $q^{\mathrm{BESS}}_{b,n,t}$ are the active and reactive power outputs at phase $n$, $SoC_{b,t}$ is the state of charge, and $E^{\mathrm{BESS}}_{b,\mathrm{nom}}$ is the nominal energy capacity.
\subsubsection{Model of Photovoltaics (PV) Output}
Behind-the-meter (BTM) PV units are assumed to operate at a constant power factor. Let $\eta_t$ denote the normalized PV output at time $t$, and let $\mathcal{B}^{\mathrm{PV}}$ denote the set of PV-equipped buses. For each $b\in\mathcal{B}^{\mathrm{PV}}$ within bus block $k$, the PV output is modeled as
\begin{subequations}\label{eq:PVcontrol}
\begin{align}
    \boldsymbol{p}^{\mathrm{PV}}_{b,t} &= \tfrac{1}{3}u^{\mathrm{BK}}_{k,t}\eta_t S^{\mathrm{PV}}_{b,\mathrm{nom}}\boldsymbol{1}_{\Phi_b},\\
    \boldsymbol{q}^{\mathrm{PV}}_{b,t} &= \boldsymbol{p}^{\mathrm{PV}}_{b,t}\tan\!\left(\varphi^{\mathrm{PV}}_{b}\right),
\end{align}
\end{subequations}
where $S^{\mathrm{PV}}_{b,\mathrm{nom}}$ is the nominal apparent power rating and $\varphi^{\mathrm{PV}}_{b}$ is the power factor angle.
\subsubsection{Model of Cold Load Pick-up (CLPU) Effect}
During BS, load demand may temporarily exceed its nominal level due to the simultaneous restoration of thermal and motor loads. This CLPU effect is modeled using a staircase approximation.

For CLs at bus $b\in\mathcal{B}^{\mathrm{CL}}_{k}$, the active and reactive demands are given by
\begin{subequations}\label{eq:clpu_cl}
\begin{align}
    \boldsymbol{p}_{b,t}^{\mathrm{CL}} = &\;\boldsymbol{p}^{\mathrm{LD}}_{b,t}\left[\textstyle\sum_{o=1}^{3}\left(\beta_{o} \Delta u^{\mathrm{BK}}_{k,t-(o-1)}\right) + u^{\mathrm{BK}}_{k,t}\right],\\
    \boldsymbol{q}_{b,t}^{\mathrm{CL}} = &\;\boldsymbol{p}_{b,t}^{\mathrm{CL}}\tan\left(\varphi^{\mathrm{CL}}_b\right),
\end{align}
\end{subequations}
where $\beta_{1}$, $\beta_{2}$, and $\beta_{3}$ are CLPU amplification coefficients, $\boldsymbol{p}^{\mathrm{LD}}_{b,t}$ is the nominal active load vector, and $\varphi^{\mathrm{CL}}_{b}$ is the power factor angle.

Similarly, for NLs at bus $b\in\mathcal{B}^{\mathrm{NL}}_{k}$,
\begin{subequations}\label{eq:clpu_nl}
\begin{align}
    \boldsymbol{p}_{b,t}^{\mathrm{NL}} = &\;\boldsymbol{p}^{\mathrm{LD}}_{b,t}\left[\textstyle\sum_{o=1}^{3}\left(\beta_{o} \Delta u^{\mathrm{NLB}}_{b,t-(o-1)}\right) + u^{\mathrm{NLB}}_{b,t}\right],\\
    \boldsymbol{q}_{b,t}^{\mathrm{NL}} = &\;\boldsymbol{p}_{b,t}^{\mathrm{NL}}\tan\left(\varphi^{\mathrm{NL}}_b\right),\\
    u^{\mathrm{NLB}}_{b,t} \le &\; u^{\mathrm{BK}}_{k,t},\\
    u^{\mathrm{NLB}}_{b,t - 1}\le &\;u^{\mathrm{NLB}}_{b,t},
\end{align}
\end{subequations}
where $u^{\mathrm{NLB}}_{b,t}$ is a binary variable indicating the energized status of the NL at bus $b$, which is dominated by the bus block's status.
\subsubsection{Model of Unbalanced Linear Power Flow}
Let $\mathcal{L}^{\mathrm{ch}}_{b}$ denote the set of child lines of bus $b$. For each bus $b\in\mathcal{B}$ with parent line $l$ at time $t$, nodal power balance is written as
\begin{subequations}\label{eq:power_flow_unbalanced}
   \begin{align}   
    \boldsymbol{\chi}_{b,t}^{\Phi_{l'}} &= \textstyle\sum_{l'\in \mathcal{L}_b^{ch}}\boldsymbol{\chi}_{l',t} - \boldsymbol{\chi}_{l,t}^{\Phi_{l'}}\\
    \boldsymbol{\chi}_{b,t} &= \boldsymbol{\chi}^{\mathrm{TG}}_{b,t} + \boldsymbol{\chi}^{\mathrm{BESS}}_{b,t} + \boldsymbol{\chi}^{\mathrm{PV}}_{b,t} - \boldsymbol{\chi}^{\mathrm{CL}}_{b,t} - \boldsymbol{\chi}^{\mathrm{NL}}_{b,t}
\end{align}
\end{subequations}
where $\boldsymbol{\chi}_{b,t}$ denotes the net active/reactive power injection at bus $b$, and $(\cdot)^{\Phi_{l'}}$ extracts the phase components associated with line $l'$.

Once the TG is available, its voltage is fixed at 1.0 p.u. For each line $l=(b,b')\in\mathcal{L}$, the sending- and receiving-end voltages satisfy
\begin{subequations}\label{eq:line_flow_unbalanced}
\begin{align}
    \boldsymbol{v}_{b',t} &\le \boldsymbol{v}_{b,t}^{\Phi_{l}}
    - 2\left(\boldsymbol{r}_{l}\boldsymbol{p}_{l,t}+\boldsymbol{x}_{l}\boldsymbol{q}_{l,t}\right)
    + \left(1-u^{\mathrm{L}}_{l,t}\right)\lceil \boldsymbol{v}_{b}^{\Phi_{l}} \rceil, \label{eq:line_flow_r}\\
    \boldsymbol{v}_{b',t} &\ge \boldsymbol{v}_{b,t}^{\Phi_{l}}
    - 2\left(\boldsymbol{r}_{l}\boldsymbol{p}_{l,t}+\boldsymbol{x}_{l}\boldsymbol{q}_{l,t}\right)
    - \left(1-u^{\mathrm{L}}_{l,t}\right)\lceil \boldsymbol{v}_{b}^{\Phi_{l}} \rceil, \label{eq:line_flow_l}
\end{align}
\end{subequations}
where $\boldsymbol{v}_{b,t}$ is the squared voltage magnitude vector at bus $b$, and $\boldsymbol{r}_{l}$ and $\boldsymbol{x}_{l}$ are the resistance- and reactance-related matrices defined in \cite{Cheng2022}.

Finally, operational security is enforced by
\begin{subequations}\label{eq:powerflowsecurity}
    \begin{align}
    -u^{\mathrm{L}}_{l,t}\boldsymbol{\chi}_{l,\max} &\le \boldsymbol{\chi}_{l,t} \le u^{\mathrm{L}}_{l,t}\boldsymbol{\chi}_{l,\max},
    \quad \forall l\in\mathcal{L}\setminus\mathcal{L}^{\mathrm{SSW}},\\
    u^{\mathrm{B}}_{b,t}\lfloor \boldsymbol{v}_{b} \rfloor &\le \boldsymbol{v}_{b,t} \le u^{\mathrm{B}}_{b,t}\lceil \boldsymbol{v}_{b} \rceil,
    \quad \forall b\in\mathcal{B}.
    \end{align}
\end{subequations}
\section{Constraint-Aware Graph Learning-Based Acceleration of SSDMGF Restoration}\label{sec:iv}
This section first introduces the graph representation used to learn synchronization patterns and the associated input features. It then presents the holistic constraint-aware STGCN, in which feasibility is resolved on the fly. Finally, the synchronization-safe restoration acceleration framework is summarized using the learned predictions.
\vspace{-1em}
\subsection{Graph Representation and Feature Construction}
The backbone graph $\mathcal{G}=\left(\mathcal{K}, \mathcal{E}^{\mathrm{SW}}\right)$ defined in Section~\ref{sec:ii} is adopted as the underlying topology for graph learning. Although the system cannot operate with all bus blocks energized and all switches closed simultaneously, $\mathcal{G}$ serves as a candidate interaction graph that bounds all feasible restoration configurations.

The time-varying availability of bus blocks and switches during restoration is not represented by changing the graph topology. Instead, it is encoded through node- and edge-level features, allowing the learning process to be performed on a fixed graph with consistent dimensions across time steps and scenarios.

Let $\mathcal{F}$ denote the set of node feature channels. At each restoration step $t$, each bus block $k$ is associated with a feature vector $\boldsymbol{X}_{t,k}\in\mathbb{R}^{|\mathcal{F}|}$, defined as
\begin{align}
    \boldsymbol{X}_{t,k}=\Big[
    &p^{\mathrm{CL}}_{t,k},\,p^{\mathrm{NL}}_{t,k},\,p^{\mathrm{PV}}_{t,k},\,u^{\mathrm{TG}}_{t},\,y^{\mathrm{dmg}}_{k},\,y^{\mathrm{BESS}}_{k},\notag\\
    &S^{\mathrm{BESS}}_{k},\,E^{\mathrm{BESS}}_{k},\,N^{\mathrm{ESW}}_{k},\,N^{\mathrm{SSW}}_{k}
    \Big],
\end{align}
where $p^{\mathrm{CL}}_{t,k}$, $p^{\mathrm{NL}}_{t,k}$, and $p^{\mathrm{PV}}_{t,k}$ denote the total CL demand, NL demand, and PV output of bus block $k$ at time $t$, respectively. $y^{\mathrm{dmg}}_{k}$ and $y^{\mathrm{BESS}}_{k}$ are binary indicators of whether bus block $k$ is damaged or equipped with a BESS. $S^{\mathrm{BESS}}_{k}$ and $E^{\mathrm{BESS}}_{k}$ denote the rated apparent power and energy capacity of the BESS, if present. $N^{\mathrm{ESW}}_{k}$ and $N^{\mathrm{SSW}}_{k}$ are the numbers of ESWs and SSWs associated with $k$, respectively.

Stacking $\boldsymbol{X}_{t,k}$ over all $t\in\mathcal{T}$ and $k\in\mathcal{K}$ yields the node feature tensor $\boldsymbol{X}\in\mathbb{R}^{|\mathcal{T}|\times|\mathcal{K}|\times|\mathcal{F}|}$, whose dimensions correspond to the time horizon, the number of bus blocks, and the number of feature channels, respectively.

The node feature vector integrates heterogeneous information, including local time-varying quantities (e.g., $p^{\mathrm{CL}}_{t,k}$ and $p^{\mathrm{PV}}_{t,k}$), local time-invariant attributes (e.g., $y^{\mathrm{dmg}}_{k}$ and $E^{\mathrm{BESS}}_{k}$), and global time-varying signals (e.g., $u^{\mathrm{TG}}_{t}$). Time-invariant and global features are broadcast along the corresponding dimensions to obtain a unified tensor representation compatible with spatio-temporal graph learning.

For each edge $e\in\mathcal{E}^{\mathrm{SW}}$, a static edge feature $y^{\mathrm{SSW}}_{e}$ is defined to indicate whether the edge is equipped with an SSW. Stacking this feature over all $e\in\mathcal{E}^{\mathrm{SW}}$ yields the edge feature vector
\begin{equation}
    \boldsymbol{E}\in\mathbb{R}^{|\mathcal{E}^{\mathrm{SW}}|}.
\end{equation}
\vspace{-3em}
\subsection{Constraint-Aware STGCN}
\subsubsection{Baseline STGCN Outputs}
The baseline STGCN is adopted as a spatio-temporal feature extractor on the backbone graph $\mathcal{G}$ to produce unconstrained predictions for synchronization-related decisions. Standard STGCN layers are omitted here for brevity, as they are well established in~\cite{Yu2017}.

Let $0$ denote the dead state of a bus block, and define the extended root set as $\mathcal{R}=\{0\}\cup\mathcal{K}^{\mathrm{BS}}$. Given the node feature tensor $\boldsymbol{X}$, the edge feature vector $\boldsymbol{E}$, and the backbone graph $\mathcal{G}$, the STGCN outputs two sets of logits:
\begin{equation}
    \left(\boldsymbol{Z}^{\mathrm{root}},\boldsymbol{Z}^{\mathrm{sync}}\right) = \mathcal{S}\left(\boldsymbol{X},\boldsymbol{E},\mathcal{G}\right)
\end{equation}
where $\boldsymbol{Z}^{\mathrm{root}}\in\mathbb{R}^{|\mathcal{T}|\times|\mathcal{K}|\times|\mathcal{R}|}$ contains node-level logits for assigning each bus block to an energized MG or the dead state, and $\boldsymbol{Z}^{\mathrm{sync}}\in\mathbb{R}^{|\mathcal{T}|\times|\mathcal{E}^{\mathrm{SSW}}|}$ contains edge-level logits for candidate SSW closures.

These logits encode learned spatio-temporal synchronization patterns, but do not guarantee feasibility with respect to synchronization safety.
\subsubsection{Constraint-Aware Feasibility Resolution Module}
Direct use of vanilla STGCN outputs is insufficient for synchronization-safe restoration, since unconstrained predictions may violate island consistency, synchronization exclusivity, or temporal causality. To address this issue, a constraint-awareness module is introduced to resolve infeasible outputs during the forward pass. The module maintains time-coupled island labels that propagate across restoration steps. At each time step, candidate synchronization actions are filtered, ranked, and selected to satisfy the synchronization-related constraints in Eqs.~\eqref{eq:systemclass}--\eqref{eq:systemmodetransition}.

The feasible root assignments $\tilde{\boldsymbol{Y}}^{\mathrm{root}}$ and SSW closures $\tilde{\boldsymbol{Y}}^{\mathrm{sync}}$ are generated on the fly by applying the operator $\Psi$ to the logits: \begin{equation} \left(\boldsymbol{H}_{t},\tilde{\boldsymbol{Y}}^{\mathrm{root}}_{t},\tilde{\boldsymbol{Y}}^{\mathrm{sync}}_{t}\right) = \Psi\left(\boldsymbol{H}_{t-1},\boldsymbol{Z}^{\mathrm{root}}_{t},\boldsymbol{Z}^{\mathrm{sync}}_{t}\right), 
\end{equation}
where $\boldsymbol{H}_{t}=\left[\ell_{t},u^{\mathrm{E}}_{t},\rho_{t}\right]$ collects the dynamic states of the feasibility-resolution module. Their roles are described below.

Let $\boldsymbol{P}^{\mathrm{root}}_{t,k}$ denote the softmax probability vector associated with $\boldsymbol{Z}^{\mathrm{root}}_{t,k}$. The root-label assignment rule for each bus block is
\begin{equation}
    \mathcal{A}\left(\boldsymbol{P}^{\mathrm{root}}_{t,k}\right)=
    \begin{cases}
    r^{\mathrm{TG}}, \qquad P^{\mathrm{root}}_{t,k,r^{\mathrm{TG}}}>\lambda,\\
    0, \qquad P^{\mathrm{root}}_{t,k,r^{\mathrm{TG}}}\leq \lambda,P^{\mathrm{root}}_{t,k,0}> \lambda,\\
    \displaystyle \arg\max_{r\in\mathcal{R}}P^{\mathrm{root}}_{t,k,r}, \qquad\text{otherwise.}
    \end{cases}
\end{equation}

Define $\ell_t(k)\in\mathcal{R}$ as the root label of bus block $k$ at time $t$. At $t=0$, the initial labels are \begin{equation} \ell_{0}(k)=\mathcal{A}\left(\boldsymbol{P}^{\mathrm{root}}_{0,k}\right),\qquad \forall k\in\mathcal{K}. 
\end{equation}

For $t>0$, root labels are inherited from the previous step until updated by accepted synchronization actions.

For each $e\in\mathcal{E}^{\mathrm{SSW}}$, define $u^{\mathrm{E}}_{t,e}\in\{0,1\}$ to indicate whether SSW $e$ is closed at time $t$. Initially, all SSW indicators are set to zero. Based on these indicators, the candidate SSW set is defined as 
\begin{equation} \mathcal{E}^{\mathrm{cand}}_{t} = \left\{ e:\boldsymbol{Z}^{\mathrm{sync}}_{t,e}>0,\; u^{\mathrm{E}}_{t-1,e}=0 \right\}. 
\end{equation}

A candidate edge is productive only if it connects two distinct islanded MGs:
\begin{align}
    \mathcal{E}^{\mathrm{prod}}_{t}=\{&e=(k,k')\in\mathcal{E}^{\mathrm{cand}}_t:\ell_{t-1}(k)\neq 0,\notag\\
    &\ell_{t-1}(k')\neq 0,\ell_{t-1}(k)\neq\ell_{t-1}(k')\}.
\end{align}

The productive set $\mathcal{E}^{\mathrm{prod}}_{t}$ is then greedily sorted in descending order of $\boldsymbol{Z}^{\mathrm{sync}}_{t,e}$. Feasible closures are selected subject to \begin{equation} \left\{\ell_{t-1}(k),\ell_{t-1}(k')\right\} \cap \left\{\ell_{t-1}(k''),\ell_{t-1}(k''')\right\} =\emptyset, \end{equation} for any distinct $(k,k'),(k'',k''')\in\mathcal{E}^{\mathrm{prod}}_{t}$. The resulting set is denoted by $\mathcal{E}^{\mathrm{prod,gree}}_{t}$.

For each SSW, accepted actions update the states as \begin{equation} \left(\tilde{Y}^{\mathrm{sync}}_{t,e},u^{\mathrm{E}}_{t,e}\right)= \begin{cases} (1,1), & e\in\mathcal{E}^{\mathrm{prod,gre}}_{t},\\ (0,u^{\mathrm{E}}_{t-1,e}), & \text{otherwise}. \end{cases} 
\end{equation}

Define $\rho_t$ as the dynamic representative map applied to the bus blocks' root labels, with $\rho_0(\ell_0)=\ell_0$. For $t>0$, $\rho_t$ is updated by each accepted $e=(k,k')\in\mathcal{E}^{\mathrm{prod,gre}}_{t}$ as \begin{subequations}
\begin{align}
    &\rho_{t}\leftarrow\mathrm{Unite}(\rho_{t-1}, \ell_{t-1}(k),\ell_{t-1}(k')),\\
    \mathrm{Unite}(\rho,i,j)\hspace{-0.2em}&=\hspace{-0.2em}
    \begin{cases}
        \rho(j)\leftarrow i, &\hspace{-1em}i\in\mathcal{K}^{\mathrm{TG}},\\
        \rho(i)\leftarrow j, &\hspace{-1em}j\in\mathcal{K}^{\mathrm{TG}},\\
        \rho[\max(i,j)]\leftarrow \min(i,j), &\hspace{-1em}\text{otherwise.}
    \end{cases}
\end{align}
\end{subequations}

The root labels are then updated as
\begin{equation}
\ell_t(k)= 
\begin{cases} \mathcal{A}\left(\boldsymbol{P}^{\mathrm{root}}_{t,k}\right), & \ell_{t-1}(k)=0,\\ \rho_t\!\left[\ell_{t-1}(k)\right], & \ell_{t-1}(k)\neq 0. 
\end{cases} 
\end{equation}

Based on the updated root labels, the feasible root assignment for each bus block is 
\begin{equation} \tilde{Y}^{\mathrm{root}}_{t,k,r}=\mathbb{1}\!\left[r=\ell_t(k)\right], 
\end{equation}
where $\mathbb{1}(\cdot)$ is the indicator function.

Although feasibility is enforced through hard logical decisions, differentiability is preserved via the straight-through estimator (STE). Hard synchronization and root-assignment decisions are used in the forward pass, while gradients are propagated through the corresponding soft outputs obtained from sigmoid ($\sigma(\cdot)$) and softmax transformations. The STE is defined as
\begin{equation}
    \mathrm{STE}(h,g) := h+\left[g-\mathrm{stopgrad}(g)\right],
\end{equation}
where
\begin{align}
    \mathrm{stopgrad}(g) &= g,\notag\\
    \frac{\partial}{\partial g}\mathrm{stopgrad}(g)&=0.\notag
\end{align}
\begin{figure*}[!t]
    \centering
    \includegraphics[width=0.8\linewidth]{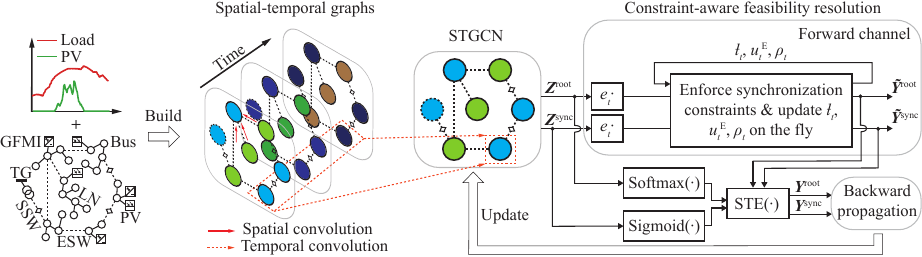}
    \vspace{-1em}
    \caption{Overall architecture of the proposed constraint-aware STGCN for synchronization-safe restoration.}
    \vspace{-1em}
    \label{fig:CA-STGCN}
\end{figure*}

The STE-wrapped feasible root assignments and SSW closures are expressed as
\begin{subequations}
    \begin{align}\boldsymbol{Y}^{\mathrm{root}}&=\mathrm{STE}\left[\tilde{\boldsymbol{Y}}^{\mathrm{root}}, \mathrm{softmax}\left(\boldsymbol{Z}^{\mathrm{root}}\right)\right],\\
        \boldsymbol{Y}^{\mathrm{sync}}&=\mathrm{STE}\left[\tilde{\boldsymbol{Y}}^{\mathrm{sync}}, \sigma\left(\boldsymbol{Z}^{\mathrm{sync}}\right)\right].
    \end{align}
\end{subequations}

Thus, the forward pass adopts the discrete feasible decisions $\tilde{\boldsymbol{Y}}^{\mathrm{root}}$ and $\tilde{\boldsymbol{Y}}^{\mathrm{sync}}$, while the backward pass approximates gradients using the corresponding soft probabilities.

The objective function of the constraint-aware STGCN is defined as
\begin{equation}
\mathcal{J} = \mathcal{J}_{\mathrm{root}} +\mathcal{J}_{\mathrm{sync}} +\gamma\mathcal{J}_{\mathrm{spar}} +\eta\mathcal{J}_{\mathrm{temp}},
\end{equation}
where $\mathcal{J}_{\mathrm{root}}$ and $\mathcal{J}_{\mathrm{sync}}$ are supervised losses on the STE outputs, and $\mathcal{J}_{\mathrm{spar}}$ and $\mathcal{J}_{\mathrm{temp}}$ are regularizers for sparsity of SSW closures and temporal smoothness of root assignments, respectively. The detailed expressions are provided in Appendix~\ref{appen:obj}.

Fig.~\ref{fig:CA-STGCN} illustrates the overall architecture of the proposed constraint-aware STGCN, including the baseline STGCN backbone and the feasibility-resolution module.
\vspace{-1em}
\subsection{Synchronization-Safe Restoration Acceleration Framework}
The objective of the learning model is not to replace the optimization-based restoration formulation, but to provide informative guidance that accelerates its solution. Accordingly, the constraint-aware STGCN is trained to predict synchronization monitor binary variables and SSW closure decision cues derived from high-quality optimization solutions. These predictions are subsequently used to construct warm-start solutions for the synchronization-safe restoration problem to release those complicated safe synchronization constraints' stress at the beginning of the solution. 







Importantly, all learning outputs are treated as advisory inputs; final restoration decisions are always validated and enforced by the optimization model to guarantee physical feasibility and synchronization safety.
\vspace{0em}
\section{Results}\label{sec:v}
This section first introduces the studied feeder and the scenario generation procedure. Next, the synchronization safety of the proposed formulation is validated through comparisons with benchmark restoration methods. Then, the synchronization-safe restoration performance under a representative winter-day scenario is presented. Finally, the effectiveness of the proposed acceleration method, based on predicted partial warm starts, is demonstrated through comparisons with alternative techniques.
\vspace{-1em}
\subsection{Simulation Setup}
\subsubsection{Feeder Description}
\begin{figure}[htbp]
    \centering
    \includegraphics[width=0.9\linewidth]{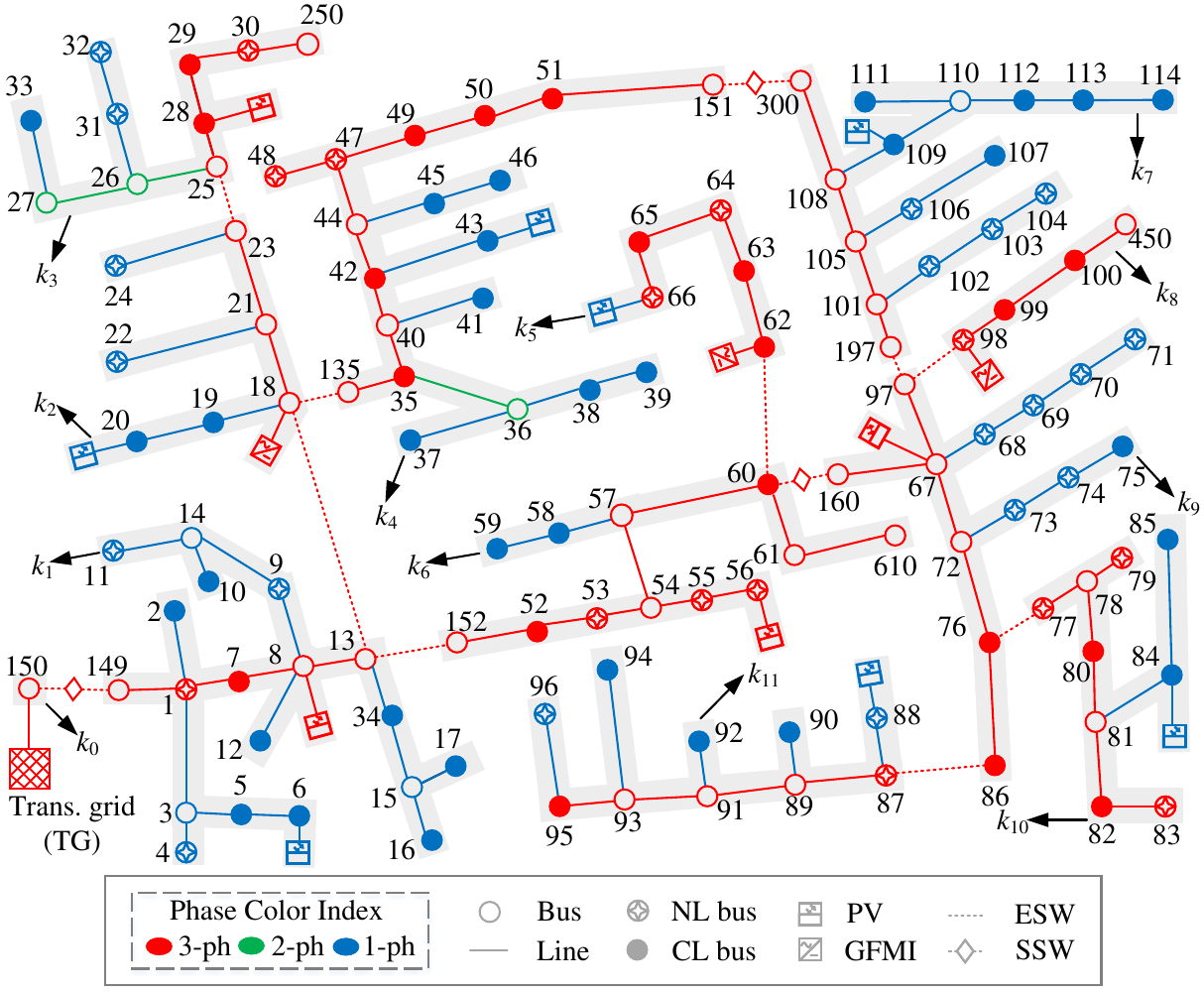}
    \vspace{-1em}
    \caption{Modified IEEE 123-node feeder with three GFMIs.}
    \vspace{-1em}
    \label{fig:testedfeeder}
\end{figure}

The test system is adapted from the three-phase unbalanced IEEE 123-node feeder, as illustrated in Fig.~\ref{fig:testedfeeder}, where colors indicate the number of phases and symbols represent different types of devices.
\vspace{-0.5em}

\begin{figure}[htbp]
    \centering
    \includegraphics[width=\linewidth]{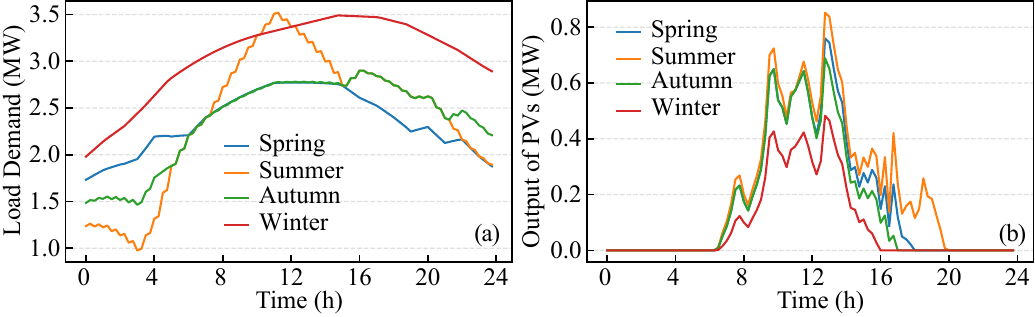}
    \vspace{-2em}
    \caption{Season-based load demand and PV output curves.}
    \vspace{-1em}
    \label{fig:loadpvscenarios}
\end{figure}

BTM distributed PV systems are deployed across the feeder, with a total capacity of $0.965$~MW, accounting for $28\%$ of the system peak demand. The power factors for load demand and PV generation are uniformly set to $0.911$ and $0.943$, respectively. The seasonal profiles of load demand and PV output are shown in Fig.~\ref{fig:loadpvscenarios}.
\vspace{-0.5em}

Three GFMI-based BESS units are installed at buses $18$, $62$, and $98$. Their rated apparent powers are $2.294$~MW, $1.283$~MW, and $2.222$~MW, with corresponding energy capacities of $3.942$~MWh, $2.471$~MWh, and $3.587$~MWh, respectively. The GFMI parameters and frequency security limits follow our previous work~\cite{Maharjan2025}. The DS is partitioned into $12$ bus blocks using $9$ ESWs and $3$ SSWs.
\subsubsection{Scenario Generation}
A diverse set of restoration scenarios is generated by varying the season, restoration starting time $t_{0}$, TG outage duration $\nu$, and the location of damaged bus blocks $k^{\mathrm{damg}}$. These scenarios provide sufficient data for training and testing the proposed constraint-aware STGCN.

Specifically, $t_{0}\in\{6,7,\dots,16\}$ (in hours), $\nu\in\{60,120,240\}$ (in minutes), and damaged bus blocks $k^{\mathrm{damg}}\in\mathcal{K}\setminus\mathcal{K}^{\mathrm{BS}}$ is considered. Across four seasons, this results in a total of $1056$ distinct restoration scenarios.
\vspace{-1em}
\subsection{Case I: Synchronization Safety Validation}
The restoration problem is evaluated under a representative winter-day scenario with $t_{0}=13$, $\nu=240$, and a damaged bus block $k_{11}$. Three MG evolution strategies are compared: the proposed SSDMGF-enabled restoration method, a rule-based restoration (RR) method, and a normal dynamic MG formation (NDMGF)-enabled method~\cite{Maharjan2025}. This comparison aims to validate the synchronization safety achieved by the proposed framework.

\begin{figure}[htbp]
    \centering
    \includegraphics[width=\linewidth]{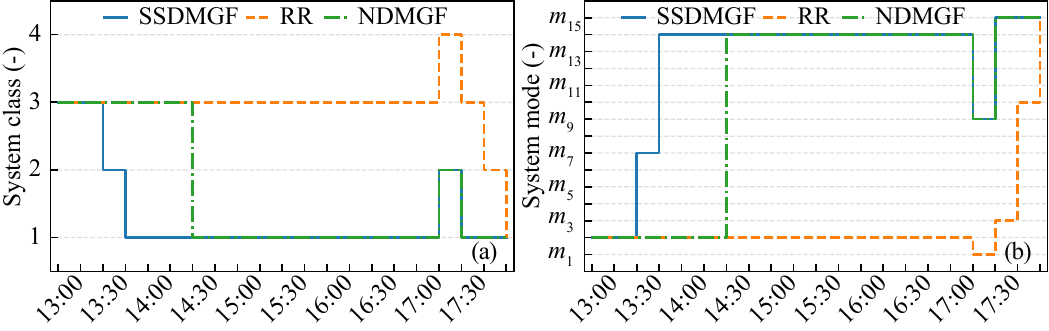}
    \vspace{-2em}
    \caption{System class and mode transitions during restoration.}
    \vspace{-1em}
    \label{fig:systemcalssandmode}
\end{figure}

For the modified IEEE 123-node system, the system class set is given by $\mathcal{C}=\{4,3,2,1\}$, derived from a total of 15 system modes (detailed in Appendix~\ref{appen:systemmode}). By indexing these modes according to their order in the appendix, the system class and mode transitions for the three methods are illustrated in Fig.~\ref{fig:systemcalssandmode}.
\vspace{-0.5em}

As shown in Fig.~\ref{fig:systemcalssandmode}(a), all three methods exhibit a generally decreasing trend in system class, except for the increase caused by TG recovery, indicating progressive MG merging during restoration. However, distinct differences can be observed. The RR method only permits MG interconnection after the TG becomes available as a reference, whereas both SSDMGF and NDMGF enable dynamic MG formation at earlier stages of the restoration.

More detailed MG interconnection processes can be inferred from Fig.~\ref{fig:systemcalssandmode}(b), together with the system mode definitions in Appendix~\ref{appen:systemmode}. Both the RR and SSDMGF methods avoid unsafe system mode transitions throughout the restoration process. In contrast, the NDMGF method, which enables dynamic MG formation without explicitly enforcing synchronization safety, introduces a risky transition at $14{:}15$, where three BS sources are merged simultaneously. Such a transition may violate synchronization constraints and lead to unsafe operating conditions.

\begin{figure}[htbp]
    \centering
    \includegraphics[width=\linewidth]{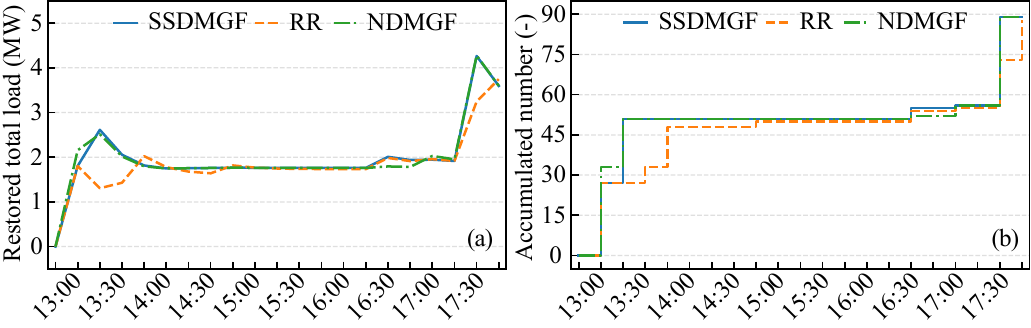}
    \vspace{-2em}
    \caption{Restored total load and accumulated number of restored customers.}
    \vspace{-1em}
    \label{fig:restoredloadandaccumulated}
\end{figure}

The corresponding restored total load and accumulated number of restored customers for the three methods are shown in Fig.~\ref{fig:restoredloadandaccumulated}. Both SSDMGF and NDMGF outperform the RR method in terms of restoration performance, achieving higher restored load and a larger number of restored customers. This improvement is mainly attributed to their ability to enable dynamic MG formation prior to TG recovery, thereby facilitating flexible sharing of BESS capacities. The NDMGF method exhibits a slight performance advantage over SSDMGF at certain stages by enabling earlier resource sharing. However, this gain comes at the cost of introducing unsafe synchronization events. In particular, the premature merging of multiple BS sources may lead to excessive inrush currents and requires additional centralized coordination to ensure secure synchronization. In contrast, the proposed SSDMGF method achieves a balanced trade-off by maintaining synchronization safety while preserving most of the performance benefits of dynamic MG formation.
\vspace{-0.5em}
\vspace{-1em}
\subsection{Case II: Synchronization-safe Restoration Performance}
The system performance achieved by the SSDMGF-enabled restoration under the same representative winter scenario is presented to illustrate the dynamic behaviors of key system components.

\begin{figure}[htbp]
    \centering
    \includegraphics[width=\linewidth]{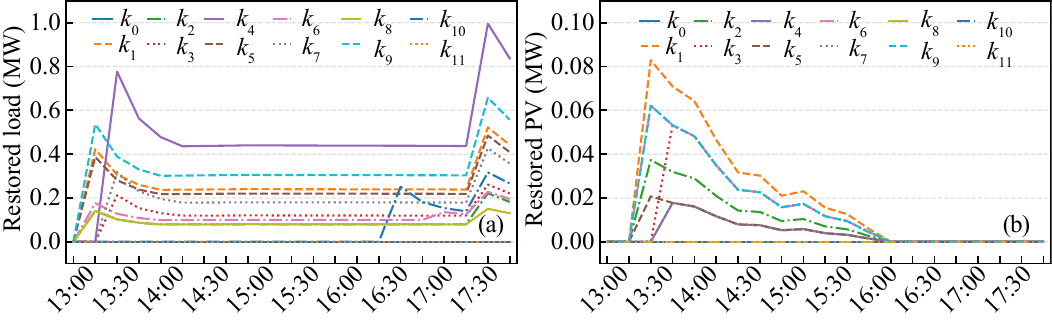}
    \vspace{-2em}
    \caption{Restored load and PV system outputs segment-by-segment.}
    \vspace{-1em}
    \label{fig:restoredloadandPVsegment}
\end{figure}

The restored load demand and distributed PV outputs at the bus-block level are shown in Fig.~\ref{fig:restoredloadandPVsegment}.
\vspace{-0.5em}

As illustrated in Fig.~\ref{fig:restoredloadandPVsegment}(a), the restored load of each bus block exhibits a clear CLPU effect upon energization. A secondary increase in load demand is observed during later stages of restoration, indicating that NLs are restored only after sufficient system capacity becomes available, particularly following TG recovery.

Fig.~\ref{fig:restoredloadandPVsegment}(b) captures the intentional activation delay of distributed PV systems, where their outputs lag behind the corresponding energization instants by one time step. This behavior is consistent with practical PV reconnection requirements and reflects the temporal coordination enforced in the restoration model.

\begin{figure}[htbp]
    \centering
    \includegraphics[width=\linewidth]{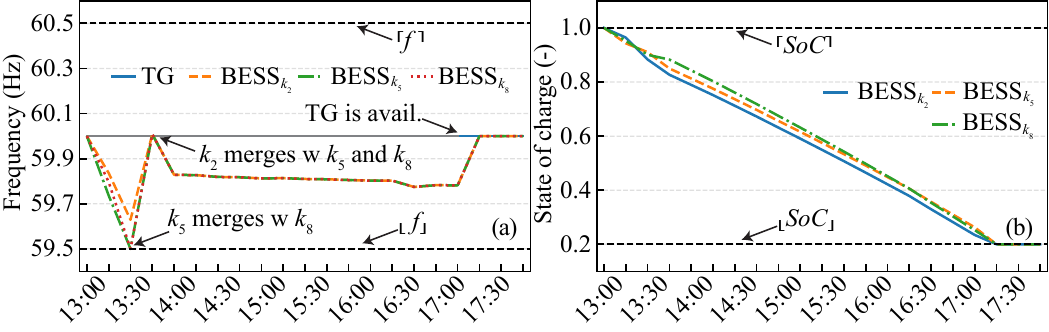}
    \vspace{-2em}
    \caption{GFMI-based BESSs' frequencies and SoCs.}
    \vspace{-1em}
    \label{fig:frequenciesandSoC}
\end{figure}

The frequencies and states of charge (SoCs) of the GFMI-based BESS units are shown in Fig.~\ref{fig:frequenciesandSoC}.
\vspace{-0.5em}

As shown in Fig.~\ref{fig:frequenciesandSoC}(a), system frequencies are consistently maintained within the prescribed safety limits. At each synchronization instant, the frequencies of the BS sources involved are well aligned, confirming that synchronization actions are executed only under feasible and secure conditions. Fig.~\ref{fig:frequenciesandSoC}(b) shows that the SoCs of all BESS units remain within the operational bounds of $[0.2, 1.0]$ p.u., indicating effective energy management throughout the restoration process.

\begin{figure}[htbp]
    \centering
    \includegraphics[width=\linewidth]{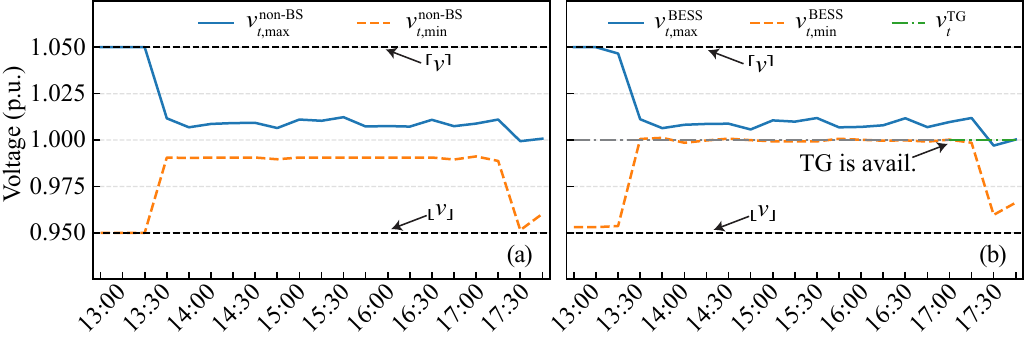}
    \vspace{-2em}
    \caption{System-wide voltage levels during the restoration.}
    \vspace{-1em}
    \label{fig:systemwidevoltage}
\end{figure}

The system-wide voltage profiles are presented in Fig.~\ref{fig:systemwidevoltage}.
\vspace{-0.5em}

As illustrated in Fig.~\ref{fig:systemwidevoltage}(a), the maximum and minimum nodal voltages of non-BS buses remain within the acceptable operational range of $[0.95, 1.05]$ p.u. throughout the restoration. In Fig.~\ref{fig:systemwidevoltage}(b), the voltages at BESS-based BS buses exhibit controlled fluctuations within safe limits, while the TG bus maintains a regulated voltage of 1.0 p.u. after reconnection. These results verify that voltage constraints are effectively enforced during the entire restoration process.
\vspace{-1em}
\subsection{Case III: Acceleration Via Partial Warm Start}
The generated scenarios are randomly partitioned into training, validation, and test sets with a ratio of $8{:}1{:}1$. The proposed constraint-aware STGCN is trained offline to generate partial warm-start solutions for the SSDMGF-enabled restoration problem.

To evaluate the effectiveness of the proposed constraint-aware warm start (CAWS), four additional initialization strategies are considered for comparison: without warm start (WWS), all-zero warm start (AZWS), random warm start (RWS), and oracle solution-based warm start (OSWS). These methods are applied to the restoration problems derived from the test scenarios, and their computational performances are recorded and compared.

The average number of infeasible warm starts, $\bar{N}_{\mathrm{IS}}$, the average number of explored nodes, $\bar{N}_{\mathrm{node}}$, and the average optimal MIP gap, $\bar{\Lambda}_{\mathrm{optm}}$, are summarized in Table~\ref{tab:computationalreco}.
\begin{table}[htbp]
\centering
\vspace{-1em}
\caption{Computational Records of Tested Warm Start Techniques}\label{tab:computationalreco}
\vspace{-0.5em}
\begin{tabular}{c c c c c}
\toprule
 Method & $\bar{N}_{\mathrm{IS}}$ & $\bar{N}_{\mathrm{node}}$ & $\bar{\Lambda}_{\mathrm{optm}}(\%)$ \\
\midrule
WWS & $/$ & 37320.55 & $4.14\times 10^{-3}$ \\
AZWS & $0$ & 27962.61 & $4.00\times 10^{-3}$ \\
RWS & $132$ & 36576.70 & $4.00\times 10^{-3}$ \\
CAWS & $0$ & 44993.31 & $4.09\times 10^{-3}$ \\
OSWS & $0$ & 47665.55 & $4.21\times 10^{-3}$ \\
\bottomrule
\end{tabular}
\end{table}

As shown in Table~\ref{tab:computationalreco}, all methods achieve comparable optimal MIP gaps, indicating that the solver converges to similar optimal solutions across different warm-start strategies. However, significant differences are observed in the number of explored nodes and warm-start feasibility.

Both AZWS, CAWS, and OSWS consistently generate feasible warm starts accepted by the solver. Specifically, AZWS provides a trivial feasible initialization without enabling MG synchronization, while CAWS produces informative and structured warm starts that guide feasible MG merging based on learned synchronization patterns. OSWS represents an idealized upper bound by directly leveraging optimal synchronization decisions.

In contrast, RWS produces infeasible warm starts in all cases, which are rejected by the solver due to violations of synchronization-related constraints. Consequently, its performance closely resembles WWS after discarding invalid initializations.

\begin{figure}[htbp]
    \centering
    \includegraphics[width=\linewidth]{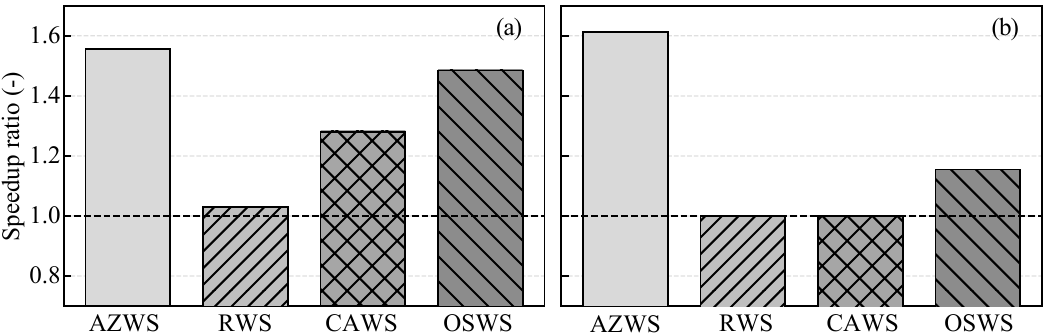}
    \vspace{-2em}
    \caption{FFS time speedup ratios of the WWS to other methods: (a) the geometric mean and (b) the median.}
    \vspace{-1em}
    \label{fig:firstfeasiblesolutiontime}
\end{figure}

\begin{figure}[htbp]
    \centering
    \includegraphics[width=\linewidth]{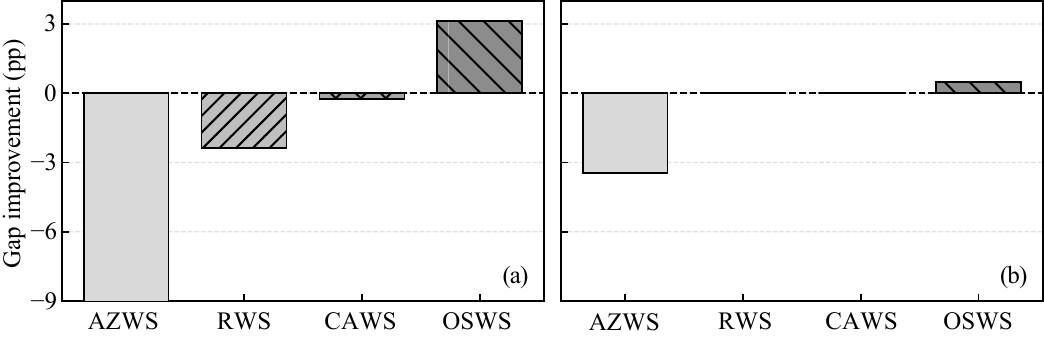}
    \vspace{-2em}
    \caption{FFS MIP gap improvement relative to the WWS: (a) the mean and (b) the median.}
    \vspace{-1em}
    \label{fig:firstfeasiblegapimprovement}
\end{figure}

The statistical comparisons of first feasible solution (FFS) time speedup ratios and MIP gap improvements relative to WWS are shown in Figs.~\ref{fig:firstfeasiblesolutiontime} and~\ref{fig:firstfeasiblegapimprovement}, respectively.
\vspace{-0.5em}

As shown in Fig.~\ref{fig:firstfeasiblesolutiontime}, both the geometric mean and median speedup ratios for AZWS, CAWS, and OSWS are significantly greater than one, indicating faster discovery of the first feasible solution compared to WWS. In contrast, the performance of RWS remains close to WWS, as infeasible warm starts are discarded and provide no effective guidance.
\vspace{-0.5em}

As illustrated in Fig.~\ref{fig:firstfeasiblegapimprovement}, AZWS results in negative MIP gap improvements, indicating that although it enables rapid identification of a feasible solution, the solution quality is relatively poor. In contrast, CAWS achieves a favorable balance between solution speed and quality, providing improved initial solutions without sacrificing feasibility.

\begin{figure}[htbp]
    \centering
    \includegraphics[width=\linewidth]{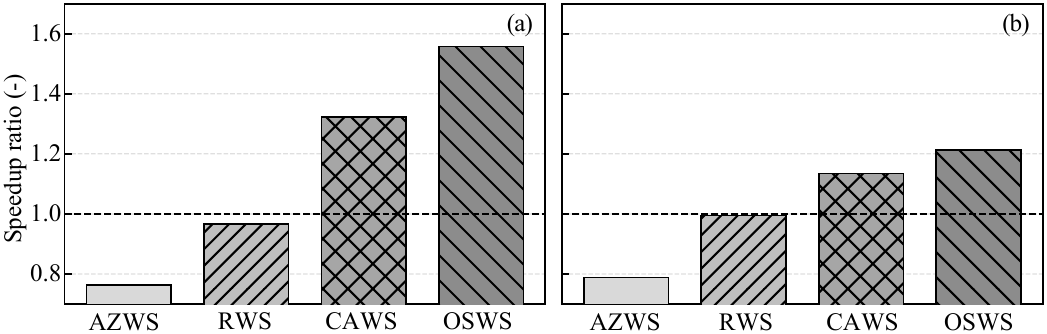}
    \vspace{-2em}
    \caption{Solution time speedup ratio of the WWS to other methods: (a) the geometric mean and (b) the median.}
    \vspace{-1em}
    \label{fig:Optimalsolutiontime}
\end{figure}

The optimal solution time speedup ratios relative to WWS are presented in Fig.~\ref{fig:Optimalsolutiontime}.
\vspace{-0.5em}

As shown in Fig.~\ref{fig:Optimalsolutiontime}, AZWS exhibits speedup ratios below one, indicating slower convergence compared to WWS due to its low-quality initialization. RWS again shows performance similar to WWS, as infeasible warm starts are discarded.

Both CAWS and OSWS achieve notable improvements in solution time by providing informative guidance on synchronization-related decisions. While OSWS, as an ideal benchmark, achieves a maximum speedup of $55.78\%$, the proposed CAWS attains a substantial speedup of $32.32\%$. It is worth noting that OSWS relies on extracting synchronization-related variables directly from the optimal solution and therefore assumes prior knowledge that is not available in practical applications. In contrast, CAWS predicts such variables via the proposed constraint-aware STGCN, making it a deployable and data-driven approach. These results demonstrate the effectiveness of the learned partial warm-start strategy in accelerating complex restoration optimization under realistic settings.
\vspace{-1em}
\section{Conclusion}\label{sec:vi}
This paper presents a synchronization-safe dynamic MG formation (SSDMGF)-enabled restoration framework for DSs, in which a constraint-aware graph learning approach is incorporated to accelerate the solution process and effectively restore loads following a blackout. To ensure safe synchronization among dynamically formed MGs with evolving boundaries, the transitions of system mode and class are explicitly restricted at each time step. Furthermore, a STE-based feasibility-resolving layer is integrated into a STGCN to generate synchronization-relaxed warm-start solutions, thereby improving computational efficiency without compromising final optimality. The key findings of this work are summarized as follows. First, the proposed SSDMGF-enabled restoration method ensures safe boundary merging by avoiding infeasible or high-risk system mode and class transitions. Second, by embedding synchronization-related constraints into the partial warm-start solutions generated by the constraint-aware STGCN, the proposed framework accelerates the solution process by 32.32\% compared with the baseline method. Future work will investigate the real-time adaptive control to further enhance the robustness of the proposed framework.
\appendix
\subsection{Constraint-Aware STGCN Objective Function Details}\label{appen:obj}
\vspace{-2em}
\begin{align}
    \mathcal{J}_{\mathrm{root}}&=\frac{1}{|\mathcal{T}|\cdot|\mathcal{K}|\cdot|\mathcal{R}|}\left\|\boldsymbol{Y}^{\mathrm{root}}-\bar{\boldsymbol{Y}}^{\mathrm{root}}\right\|_{1},\tag{A.1}\\
    \mathcal{J}_{\mathrm{sync}}&=\frac{1}{|\mathcal{T}|\cdot|\mathcal{E}^{\mathrm{SSW}}|}\left\|\boldsymbol{Y}^{\mathrm{sync}}-\bar{\boldsymbol{Y}}^{\mathrm{sync}}\right\|_{1},\tag{A.2}\\
    \mathcal{J}_{\mathrm{spar}}&=\frac{1}{|\mathcal{T}|\cdot|\mathcal{E}^{\mathrm{SSW}}|}\textstyle\sum_{t,e}\boldsymbol{Y}^{\mathrm{sync}}_{t,e},\tag{A.3}\\
    \mathcal{J}_{\mathrm{temp}}&=\frac{1}{(|\mathcal{T}| - 1)\cdot|\mathcal{K}|\cdot|\mathcal{R}|}\textstyle\sum_{t}\left\|\boldsymbol{Y}^{\mathrm{root}}_{t}-\boldsymbol{Y}^{\mathrm{root}}_{t-1}\right\|_{1}.\tag{A.4}
\end{align}
\subsection{System Modes for The Revised IEEE 123-Node Feeder}\label{appen:systemmode}
\vspace{-2em}
\begin{align}
\mathcal{M}_4 = \big\{&\{\{k_0\},\{k_2\},\{k_5\},\{k_8\}\}\big\},\tag{B.1}\\
\mathcal{M}_3 = \big\{&\{\{k_2\},\{k_5\},\{k_8\}\},
\{\{k_0,k_2\},k_5,k_8\},\notag\\
&\{\{k_0,k_5\},k_2,k_8\}, \{k_0,\{k_2,k_8\},k_5\},\notag\\
&\{k_0,k_2,\{k_5,k_8\}\}\big\},\tag{B.2}\\
\mathcal{M}_2 = \big\{&\{\{k_2\},\{k_5,k_8\}\},
\{\{k_2,k_8\},\{k_5\}\},\notag\\
&\{\{k_0\},\{k_2,k_5,k_8\}\},\{\{k_0,k_2,k_8\},k_5\},\notag\\
&\{\{k_0,k_5,k_8\},k_2\},
\{\{k_0,k_2\},\{k_5,k_8\}\}, \notag\\
&\{\{k_0,k_5\},\{k_2,k_8\}\}\big\},\tag{B.3}\\
    \mathcal{M}_1 = \big\{&\{k_2,k_5,k_8\},\{k_0,k_2,k_5,k_8\}\big\}.\tag{B.4}
\end{align}
\ifCLASSOPTIONcaptionsoff
  \newpage
\fi
\bibliographystyle{IEEEtran}
\bibliography{references.bib}
\end{document}